# Real or Fake? User Behavior and Attitudes Related to Determining the Veracity of Social Media Posts

**Linda Plotnick**
New Jersey Institute of Technology
Linda.Plotnick@gmail.com

**Starr Roxanne Hiltz**
New Jersey Institute of Technology
Roxanne.hiltx@gmail.com

**Sukeshini Grandhi**
Eastern Connecticut State University
grandhis@easternct.edu

**Julie Dugdale**
University Grenoble Alps
Grenoble Informatics Lab, France
JulieDugdale@imag.fr

**ABSTRACT**

Citizens and Emergency Managers need to be able to distinguish "fake" (untrue) news posts from real news posts on social media during disasters. This paper is based on an online survey conducted in 2018 that produced 341 responses from invitations distributed via email and through Facebook. It explores to what extent and how citizens generally assess whether postings are "true" or "fake," and describes indicators of the trustworthiness of content that users would like. The mean response on a semantic differential scale measuring how frequently users attempt to verify the news trustworthiness (a scale from 1-never to 5-always) was 3.37. The most frequent message characteristics citizens' use are grammar and the trustworthiness of the sender. Most respondents would find an indicator of trustworthiness helpful, with the most popular choice being a colored graphic. Limitations and implications for assessments of trustworthiness during disasters are discussed.

**INTRODUCTION**

Most of the studies on social media use during crises focus on Emergency Managers (EMs) in governmental or NGO organizations. For example, Reuter et al. (2018) found that papers on social media presented at ISCRAM meetings through 2017 were "mostly written with the aim to facilitate the work of first responders and emergency managers." The trustworthiness of posts from the public is a major issue for EMs in terms of its limiting the perceived usefulness of social media during a crisis (e.g., Hiltz et al, 2014). However, citizens often use social media (SM) during disasters to inform one another of conditions and events, and to coordinate citizen responses in their communities, which can increase resiliency. They need to be able to tell "real" from "fake" news posts. Beyond citizen to citizen information sharing, response agencies and NGOs are monitoring citizen postings to inform aspects of their decision making. If these postings are to provide real value for situational awareness, they too must be able to distinguish "real" from "fake" news. So, this issue is pervasive and affects all participants in crisis SM communications. In this paper, we focus on citizen assessment of the "truth" of social media posts. It is exploratory research with an emphasis on describing the current thinking and self-reported behavior of users of social media related to what is real and what is "fake," and what software enhancements they might use to help them tell the difference.

Fake news in social media refers to video, audio, and/or text content that spreads false information [Ordway, 2017]. This paper presents a study of the use of social media and behavior related to "fake news" conducted in early 2018 in the U.S. and examines general news that is not specifically related to disaster situations. However, the general behavior of SM users in this regard is relevant to crisis situations; it is likely that the user behavior patterns followed in everyday life also hold even more for crisis situations as people are eager to share and to share quickly. Although not directly related to fake news, Hughes and Palen in a study concerning Twitter, suggest that more broadcast-based information sharing activities happen during crisis events [Hughes and Palen, 2009]. Furthermore, Gupta and colleagues showed that social media is subject to fake news in emergency situations, causing panic and chaos [Gupta et al. 2013]. Nevertheless, the assumption that "everyday" behavior related to assessments of the trustworthiness of posts carries over to crisis situations needs further research.

As exploratory research, we began with research questions rather than theories and hypotheses about what we might find. The research questions addressed in this paper are:

RQ1: To what extent and how do citizen users of social media assess whether postings are "true" or "fake?"

RQ2: What kinds of indications of the trustworthiness of postings would users like to have?

This paper begins with a brief review of the use of social media by citizens during disasters and of the concerns about "fake news." A description of the online survey conducted and of the participants is followed by results related to the research questions and a discussion of their implications for disaster response and social media design features.

**LITERATURE REVIEW**

**Citizens' Use of Social Media During Crisis**

The most frequent use of social media by citizens during disasters is to search for (and post) relevant information about an evolving situation, as it affects them and their friends and families in a specific location. After the terrorist attacks of 11 September 2001, before the era of Facebook, wikis created by citizens were used to collect information on missing people (Palen & Liu, 2007). An early study of "back channel" information sharing among citizens (which do not go through or involve directly, official EMs) was of the 2007 Southern California Wildfires, which lasted for 19 days, destroyed about 1500 homes, burned over 500,000 acres, and caused massive evacuations (Sutton et al, 2008). California, like many areas, has an official and hierarchical command and control system (Incident Command System or ICS) that is supposed to oversee the one-directional communication of disaster information from the formal government agencies to the media, and thence to the public. The "back channel" is (unofficial) response activity, among the victims, observers, and responders themselves. As one San Diego County resident explained:

*The only way we all have to get good information here is for those who have it to share it. We relied on others to give us updates when they had info and we do the same for others.* (Sutton et al., 2008)

A particularly striking quote is,

*"What we learned… is that there is no 'they.' 'They' won't tell us if there is danger, 'they' aren't coming to help, and 'they' won't correct bad information. We (regular folks) have to do that amongst ourselves." (Sutton et al., 2008, p. 627).*

Besides general information exchange as a disaster unfolds, social media are also used by citizens to organize and cooperate various types of actions. Often official organizations cannot meet the demand for first responders during a large-scale disaster, due both to limited personnel and problems with damaged or inaccessible equipment. Informal citizen-led rescue groups often form during major disasters, such as Hurricane Harvey in the Greater Houston area in 2017. Citizens used diverse apps and SM related platforms to coordinate and carry out rescues (Smith et al., 2018). Likewise, in anticipation of Hurricane Sandy, online pet lovers established a Facebook page to cooperate and organize their work to assist pets displaced or abandoned as a result of the storm (White, Palen, & Anderson, 2014).

In all of these disaster response social media uses, credibility of the postings is an issue. Users may post and share untrue information either because they mistakenly believe to it to be true, or because they deliberately want to spread fear and misinformation. An example is a post of a picture supposedly of a shark swimming down a New York street during Hurricane Sandy. In studies that included Hurricane Katrina in 2005 and the 2010 volcano Eyjafjallajökull in Iceland, Endsley et al. (2014) studied how different factors (strength of social ties and sources of crisis information) affect perception of the credibility of crisis information about natural disasters. They found that the perceived credibility of SM information by users was less than that of printed, official online or televised news, or information from family, relatives or friends. In order for information to be considered "actionable," it must first be considered trustworthy and not "fake news."

**Fake News and Echo Chambers**

A broad objective of this research project is to investigate the human practices and perceptions behind information seeking and opinion formation in the context of fake news as well as "echo chambers." Echo chambers in social media refer to a relatively homogenous social space that reinforces/echoes beliefs and views within one's likeminded network while omitting or censoring opposing beliefs and views from other individuals or media sources (Wohn & Bowe, 2016). It is widely accepted that fake news and echo chambers increase the risk of misinformed personal, social and political decision-making. While factors such as social media design and underlying algorithms have been shown to influence formation of echo chambers (Pariser, 2011), other research has suggested that human cognition and biases result in people accepting and making decisions based on fake news and ideas reinforced by echo chambers (Pentland, 2013). Yet others suggest that this is very much a result of people's personal choice and behavior (Hosanagar, 2016; Baranuik, 2016). People seek resonance, and hence become friends with and interact with, likeminded people while "unfriending" those who aren't. That is, they seek information and news spread by those within their echo chambers without independently conducting due diligence. Our goal is to understand how to design social media technologies that can help people detect fake news and mitigate its negative effects. This paper represents a preliminary analysis of some of the data relevant to this objective, as it relates to citizen use of social media during crises. We will look at the extent to which people try to obtain views from categories of people or sources of information that are "different" from their own, rather than communicating within homogeneous echo chambers; this is way of assessing the tendency to believe "fake news" within their social media networks.

False rumors and misinformation often persist by being transformed and resurfaced on partisan news sites that repackage the rumors as "news"; readers then share this "news' with their Twitter followers or on other social media (Shin, Jian, Driscoll, & Bar, 2018). In a recent study (Barthel et al., 2016), 64% of 1002 US adults surveyed believed that fake news stories cause "a great deal of confusion" regarding basic facts in current issues and events; 78% say they often or sometimes see inaccurate information; and 23% say that they have at sometimes shared fake news. Understanding how and why individuals adopt such behavior can help us design features and mechanisms within social media to minimize such actions and mitigate their effects. This in turn can help people to make more informed and unbiased decisions in all walks of life.

In the realm of crisis events related to terrorism, Starbird et al (2014) conducted a study of fake news following the Boston Marathon bombing in the spring of 2013. As information spread via social media, it was filled with rumors (unsubstantiated information), and many of these rumors contained misinformation. Their exploratory research examined three rumors, later demonstrated to be false, that circulated on Twitter in the aftermath of the bombings. Their findings suggest that corrections to the misinformation emerge but are muted compared with the propagation of the misinformation.

**Tools and Practices for Filtering and Assessing Social Media**

There are now scores of studies in the literature that describe possible software enhancements and systems that could improve the usability and usefulness of social media for disaster management. See Imran et al. (2015) for a complete survey of these technologies. Building on this research, Plotnick and Hiltz (2018) conducted a survey of U.S. county level emergency managers that included their perceptions of the usefulness of technological enhancements in existence or currently being developed. The idea behind some of these tools, as well as practices described in other literature on techniques for assessing the trustworthiness and validity of social media posts, were used to develop questions for the survey described in this paper.

**METHOD**

**Data Collection**

In spring 2018 we deployed an online survey using SurveyMonkey to a university community (students and employees of a liberal arts institution in New England) and the general public through social media (Facebook). After the Institutional Review Board, the project procedures were approved, and participants were recruited through snowball sampling method with the authors requesting people to share the survey link widely with family, friends and colleagues. Participants were not offered any compensation for their time. However, for every 25 completed surveys, up to a maximum of 500, a random drawing took place offering a $25 gift. After data cleaning, the total number of usable responses included 179 university members and 198 external respondents.

**Survey Instrument**

The survey included several open-ended text questions as well as structured questions using nominal, ordinal, and interval measures to explore people's practices of assessing trustworthiness of social media posts as well as people's sharing practices. Our survey included modified questions from Morris et al (2012), in particular questions focused on the posting characteristics used by respondents to assess trustworthiness, as well as our own items culled from previous research. The modified questions from Morris et al. (2012) were used with permission from the authors (Morris et al. 2012) and had been deployed in a Microsoft survey. The final set of items for assessing trustworthiness of a post is shown in Table 4 below. Appendix 1 shows the question near the beginning of the survey defining social media and listing systems that we specifically wanted respondents to bear in mind when answering other questions. In this paper however, we report specifically on the following questions asked in the survey. In deciding how trustworthy information you receive from *social media* is, how often do you consider the following factors?

- Are there any types of news / information you are most likely to try to verify for trustworthiness on social media?
- If a social media allows users to rate the trustworthiness of news sources, how willing are you to take the time to rate various news sources?
- To what extent do you try to verify the trustworthiness of news from social media or other sources on a regular basis?
- If you could view a trustworthiness indicator for the information you are reading on social media, how would you like the indicator to look?
- Would seeing an indicator of trustworthiness influence how much you trust the news/information?
- To what extent do you consider the trustworthiness of news/information before you share it on social media?
- To get a different point of view, how often have you followed, friended, or otherwise connected to a person or source with whom you would probably disagree?

**DATA ANALYSIS AND RESULTS**

**Characteristics of Respondents**

We obtained usable responses from 341 people. Unusable responses, which we did not use in the analysis, include those that just responded to the Informed Consent but didn't respond to questions. Note, however, that some respondents did not answer all questions so the N for any one test may not reflect responses from all of our subjects. Our respondents are diverse. Roughly half (53%) are full or part-time students, while the remaining 47% are not in school. Despite that distribution, 77% (N=264) are employed which suggests that many of our respondents carry the responsibilities of both employment and academic work. While the majority of our respondents are female (58%, N=196), the gender of the respondents is nearly equally distributed with 41% (141) male and 1% (4) identifying as "other". The distribution of respondents by age is shown below in Table 1:

| AGE GROUP (in years) | FREQUENCY | Percentage |
|---|---|---|
| 18-24 | 138 | 40.46 |
| 25-34 | 61 | 17.88 |
| 35-44 | 38 | 11.14 |
| 45-54 | 48 | 14.07 |
| 55-64 | 34 | 10.00 |
| 65+ | 22 | 6.45 |
| N | 341 | 100.00 |

**Table 1. Distribution of respondents by age**

Tables 2 and 3 below show the distribution of respondents by highest academic degree earned and by reported ethnicity:

| Highest Degree | Frequency | Percentage |
|---|---|---|
| High School | 101 | 29.62 |
| Associates | 35 | 10.26 |
| Trade School | 2 | 0.59 |
| Bachelor | 64 | 18.77 |
| Master | 59 | 17.30 |
| Doctorate | 71 | 20.82 |
| Other | 9 | 2.64 |
| **N** | **341** | **100.00** |

Table 2. Distribution of respondents by highest academic degree earned

| Race/Ethnicity | Frequency | Percentage |
|---|---|---|
| American Indian/ Alaskan Native | 3 | 0.9 |
| Asian/ Pacific Islander | 41 | 12.01 |
| Black or African American | 21 | 6.16 |
| Hispanic | 18 | 5.27 |
| White/ Caucasian | 232 | 68.04 |
| Prefer not to answer | 9 | 2.64 |
| Multiple Ethnicity/ Other | 17 | 4.98 |
| **N** | **341** | **100** |

Table 3. Distribution of respondents by ethnicity

Of the 334 respondents who answered the question on what state or country they are from, 87% (289) are from the U.S. and the remaining 13% (45) hail from countries around the world. Of the U.S. respondents, 60% (173) reside in the northeastern state in which one of the authors works and thus was a focus for deploying the survey.

Thus, while most of our respondents are Caucasian, young, and are employed, they are diverse by other measures of respondent characteristics. Important to note is that 95% (324) of the respondents use social media (5% do not). We suggest that may be a result of the sample being primarily young and the ubiquitous nature of social media, and of course the fact that Facebook and other social media were a prime means for recruiting respondents.

**Methodology for Analysis**

Our survey contained nominal, ordinal, and interval data. Frequencies were taken and reported for the nominal items. Tests of normality (Kolmogorov-Smirnov) were done on the other data and it was found that our data are not normally distributed. Therefore, we used non-parametric tests (e.g. Kruskal-Wallis, Mann Whitney U) for our statistical analysis. Kruskal-Wallis is a non-parametric "version" of ANOVA but, unlike ANOVA, cannot tell you where differences in means are – just that there are significant differences. ANOVA requires normally distributed data because with non-normally distributed data it is not as sensitive (robust). That is, with non-normally distributed data it will not give a result of significance where there is none but might miss a significant relationship. Therefore, in our tests, if Kruskal-Wallis test found that there are significant differences, we then ran ANOVA with Tukey's post hoc tests to try to determine where that difference is.

**Results and Discussion**

We first present the overall results for the questions of interest for this paper, and then look at whether or not results varied by characteristics of the respondents.

*Current Techniques to Assess Trustworthiness*

We asked the respondents to assess how often they consider various factors to determine trustworthiness of data they receive from social media. The question was a semantic differential from 1 (never) to 5 (always) with an option for "don't know". For this analysis, we did not include "don't know" responses. Below is a sorted table of results from highest mean (most used) to lowest).

| Characteristic | N | Mean | Standard Deviation | Skewness |
|---|---|---|---|---|
| Grammar | 269 | 4.08 | 1.17 | -1.01 |
| Sender is trustworthy | 271 | 4.04 | 1.13 | -1.15 |
| Content mentions the source of information | 271 | 3.94 | 1.09 | -0.95 |
| Sender has verified expertise | 266 | 3.88 | 1.14 | -0.86 |
| Sender is affiliated with a trusted organization | 272 | 3.80 | 1.17 | -0.81 |
| Know sender | 274 | 3.77 | 1.19 | -0.68 |
| Contains a URL | 255 | 3.55 | 1.30 | -0.48 |
| Content of responses to the post | 271 | 3.01 | 1.29 | -0.07 |
| Several other similar posts have been shared | 266 | 2.94 | 1.24 | 0.01 |
| The type of profile picture sender has | 264 | 2.86 | 1.45 | 0.13 |
| Sender's location is relevant to the topic | 252 | 2.81 | 1.28 | 0.07 |
| Sender is a frequent poster | 253 | 2.67 | 1.40 | 0.29 |
| Position of the post in search results | 245 | 2.58 | 1.36 | 0.36 |
| Sender's location | 240 | 2.42 | 1.32 | 0.49 |
| Sender is an infrequent poster | 244 | 2.41 | 1.36 | 0.62 |

**Table 4. Characteristics considered when assessing trustworthiness of information**

The results suggest that there is a wide range of responses to characteristics that our respondents consider to be indicators of trustworthiness of social media posts. While the grammar and trustworthiness of the sender are deemed as important, the location of the sender, for example, is not a critical characteristic. That the trustworthiness of the sender is important is not a surprise. That the grammar of the post is equally important is somewhat surprising. We postulate that grammar suggests other characteristics such as the sender being informed, discerning, careful in creating the post, etc. Additionally, it may be misleading to rely on grammar as an indicator as during the stress of crisis, grammar may suffer. It would be interesting in the future to explore this.

*Frequency of verifying trustworthiness of social media data*

Related to the above, we asked in a semantic differential question (from 1-never to 5-always) how frequently the respondent does try to verify the trustworthiness of news from social media or other sources. The mean response was 3.37 (N=262, std. dev. = 1.08, skewness -.305). In another semantic differential question (from 1-never to 5-frequently) the respondents were asked to what extent they consider the trustworthiness of news/ information before sharing it on social media. The mean response for sharing news (μ=4.15, std. dev. = 1.14, skewness = -1.26) was significantly higher than for consuming news (Mann Whitney U test results of z=-2.3, p=.022). These results suggest that the respondents are more cautious about the social media posts they share than content for only their own consumption.

Given that respondents are cautious about the social media posts they share which suggests (with other results reported here) that they care about trustworthiness, we wonder how willing they would be to devote time to rate trustworthiness of news sources. Knowing this can give an indication of how important trustworthiness is to them. Our survey asked if how willing our respondents were to take the time to rate various news sources (semantic differential from 1 not willing to 5 very willing). The mean response was 3.30 (std. dev. = 1.26, skewness = -.312). As the scale ranged from 1 to 5, it is designed so that responses below the midpoint of the scale are not very willing (with the degree of reluctance dependent upon how close to the midpoint the response is) and those responses at the midpoint and above are increasingly willing. Thus, the results suggest that respondents are willing, overall, but perhaps not enthusiastically so.

Trustworthiness is, in some sense, a subjective judgment. We have seen that our respondents do care about the trustworthiness of the content they read and share on social media. We have also seen that the respondents believe some of the characteristics of the posts (Table 4) indicate levels of trustworthiness. However, we do not know if the respondents are open to learning about and perhaps trusting content that is contrary to their current beliefs. That is, how confined are the respondents to their echo chambers? This is a big research question that is too broad to be explored in depth in our study. However, to glimpse into possible answers, we asked "To get a different point of view, how often have you followed, friended, or otherwise connected to a person or source with whom you knew you would probably disagree?" The question asked for responses from 1-never to 5-many times. As above, this semantic differential scale was designed to be a continuous rating from the left endpoint (never) to the right endpoint (many times). The results ($\mu$ = 2.40, std. dev. = 1.21, skewness = .44) suggest that respondents are reluctant to explore ideas outside of their echo chambers. This finding confirms what has been suggested in the literature, that people tend to connect with those with similar values and views (Hsieh et. al, 2012; Munson and Resnick, 2010; Wang and Mark, 2017).

*Perceptions of usefulness of possible indicators of trustworthiness*

We also explored how respondents would feel about an indicator of trustworthiness attached to content. The frequencies of respondents who would like a proposed indicator are shown in Table 5. Note that respondents were free to choose as many indicators as they wished.

| No indicator | A number from 0 to 100 | Colored graphic | Dial with numbers from 0 to 100 | Sliding scale from 0 to 100 | Flashing stop light |
|---|---|---|---|---|---|
| 36 | 82 | 102 | 40 | 61 | 60 |

**Table 5. Preferences for indicators of trustworthiness**

The results indicate that the respondents, by and large, would find an indicator helpful, and that display in the form of a colored graphic is the most popular choice.

We also asked the respondents whether seeing an indicator of trustworthiness would influence how much they would trust the news/information. The results, as shown below, are less clear, although one can note that few (11%) believed that under no circumstances would there be an influence. That there were so many (almost half) who responded that "it depends" is of interest; further data collection and analysis could lead to insights as to what it depends upon.

| **Response** | **N** | **Percentage** |
|---|---|---|
| It depends | 117 | 45.5 |
| Yes | 94 | 36.6 |
| No | 28 | 10.9 |
| Not sure | 18 | 7.0 |
| Total | 257 | 100.0 |

**Table 6. Perceptions of influence of indicator on belief of trustworthiness**

*Differences in willingness to rate trustworthiness between groups of respondents*

To what extent are our results generalizable across different demographic groups? While the results in Table 6 suggest that respondents overall are willing to take the time to rate the trustworthiness of news content if they are given the opportunity, we were curious to see if respondent characteristics affected their willingness. For example, are younger respondents more, less, or equally willing than older respondents? To explore this, we used Kruskal-Wallis (nonparametric) tests to compare groups based on some of the characteristics data we collected.

Kruskal-Wallis tests did not find a significant difference in the willingness to take the time to rate content by age ($H = 3.74$, $p=.588$), by gender ($H = 2.01$, $p = .367$), by whether or not the respondent is a student ($H = .87$, $p = .351$), by whether or not the respondent is employed ($H = .29$, $p = .592$), or the highest academic degree obtained ($H=4.11$, $p = .662$). That is, the demographic characteristics of the respondent do not appear to affect their willingness to take time to rate the trustworthiness of news content on social media. Next, we looked at the differences by characteristics of respondents of the extent respondents try to verify the trustworthiness of news.

*Differences in the extent respondents try to verify the trustworthiness of news between respondent groups*

We then ran Kruskal-Wallis tests to compare how respondents grouped by the same demographics rated their willingness to try to verify the trustworthiness of news. The tests did not find any significant differences when the respondents were grouped by age ($H = 8.24$, $p = .144$), nor when grouped by gender ($H = .05$, $p = .976$), nor whether or not the respondent is a student ($H = 2.05$, $p = .152$), nor whether or not the respondent is employed ($H = .01$, $p = .925$). However, when the respondents were grouped by highest academic degree earned, the Kruskal-Wallis test did find that there is a significant difference between groups ($H = 17.42$, $p = .008$). As noted above, Kruskal-Wallis does not show between which group(s) there are significant differences, so we then ran an ANOVA with Tukey's post hoc tests. The results suggest that there is a significant difference

The ANOVA with Tukey's post hoc tests suggest that there is a significant difference between the associate's degree group and the "other degree" groups such that the associates degree group rated their willingness higher; and that there is a significant difference between the high school diploma group and the associates degree group such that, again, the associates degree group rated their willingness to try to verify trustworthiness higher. No other significant differences were found. However, it is of note that not only were the associates, bachelor's degree, master's degree, and doctoral degree groups not statistically different from one another in their ratings but the significance of $p = 1$ suggests they were the same. In addition, note that only 9 respondents said they had some "other" type of degree (Table 2) so the statistical difference is not very meaningful. We posit that those respondents who have post-high school degrees (e.g. those with associates degrees and above) spent time in academia learning to research topics and so are more comfortable with, and willing to, research/verify the trustworthiness of news content.

*Differences in the frequency respondents connect with others to gain a different point of view*

Although the findings reported above suggest that respondents are reluctant to go outside of their echo chamber to hear another point of view, we explored whether this is ubiquitously true or differs by demographics by running Kruskal-Wallis tests.

The tests did not find any statistically significant differences by age ($H = 1.84$, $p = .871$), gender ($H = 3.26$, $p = .206$), whether or not the respondent is a student ($H = .54$, $p = .463$), or the highest degree earned by the respondent ($H = 3.67$, $p = .721$). However, the Kruskal-Wallis test did detect significant differences for the grouping of respondents by whether or not they are employed ($H = 4.13$, $p = .042$). Because there are only two groups, it is not necessary to run ANOVA in this case. We ran a Mann-Whitney U Test that found a statistically significant difference ($z = -2.033$, $p = .042$) such that those respondents not employed (mean rank = 129.55) are more willing to go outside of their echo chamber than are those who are employed (mean rank = 109.26). This result is a conundrum. It is, however, possible that those who work spend most of their time with like-minded people (echo chamber) and therefore do not have the opportunities those who do not work have to be exposed to different points of views. Seeking alternate points of view may cause discomfort for the employed respondent.

In summary, for the majority of demographic characteristics that we collected data for, there are no statistically significant differences between groups when the responses to the variables of interest are compared.

**LIMITATIONS AND FUTURE RESEARCH**

As with any research project, our study has some limitations. First, the questions were about general behavior and attitudes related to discerning "fake news" postings and did not explicitly ask about how this might change during a disaster. Secondly, we were unable to obtain a random sample of users of social media and relied, instead, on a snowball convenience sample. Our respondents did not have the level of diversity that would be ideal – most were Caucasian, young, and employed. However, the fact that there were few significant differences in responses related to age, gender, educational level, or whether or not the respondent was a student, suggests that the results would be similar for a more representative sample of social media users. Third, we employed a single instrument (survey) and thus were unable to achieve triangulation. Nonetheless we were able to gain insights into the issue of fake news that not only inform but can provide suggestions for future research efforts.

In future research, we will endeavor to use sampling methods to have better representation of the population affected by fake news. We will use other research instruments, such as interviews and perhaps focus groups. Some questions should focus specifically on assessments of trustworthiness during disasters, not just on everyday use of social media. An element of scenario driven questions, using a disaster scenario such as perhaps a shooting on a local university or high school campus, would help us refine the focus of our investigation. And of course, future research should aim at theory building and testing, not just descriptive analysis. There is much more to learn, and we intend to continue to pursue this investigation.

**SUMMARY AND CONCLUSION**

The issue of fake news has been trending lately. Nowhere is it more important than during crisis. The public relies on news, including in social media, to inform them of what steps to take to be safe and proactive during crisis. This study provides insights that can help us understand better the perceptions of fake news and how to better design technologies to promote assessment of the trustworthiness of news and other information. If users can assess the veracity of social media posts related to a crisis, this can add to situation awareness, give them confidence in suggested actions to maximize their safety and minimize their losses, and thus contribute to resilience.

Our findings suggest that the trustworthiness of social media and other sources of information is important to the public. This is particularly evident in our finding that our respondents are less cautious about trustworthiness when the news is for personal consumption than when they are sharing the news. We also found that, in general, respondents are reluctant to go outside of their echo chambers to get a different point of view. They are concerned about spreading rumors and take more care when disseminating content than when using it for their own purposes. It is also shown in the high level at which participants would like to see an indicator of trustworthiness to help them assess content trustworthiness that trustworthiness is a concern. The finding that users would most like to see an indicator of the trustworthiness in the form of a colored graphic or a number from 0 to 100 is of practical importance for designers for systems like Facebook who are working towards providing such information to users.

Especially during crises, when time is limited and indicators such as the grammatical correctness of posts may not be as reliable as during "ordinary" times of less stress, some sort of graphical indicator of the computed trustworthiness of posts on the topic of the crisis appears to be a tool that would be very useful for citizens (as well as emergency managers). Such an indicator could be constructed from a combination of information about the poster obtained from a system's data (e.g., how many followers or friends; frequency of posting in recent past; sender's location) and an automated sentiment analysis of replies to the post that looks for agreement vs. disagreement with the post (see, e.g., Halse, Binda, and Weirman, 2018).

Finally, in our study, there were few differences in responses of groups determined by demographics. This has implications for design of social media platforms. If this finding is also shown in future research, there may be "one size fits all" solutions to technological design of indicators and applications targeted at helping users assess trustworthiness.

There is much to explore, but our study provides a good foundation for moving forward in this area. As noted above, subsequent studies should explicitly focus on assessments of trustworthiness of social media posts during crises and should begin to build and test a theoretical model of the determinants of whether and how users engage in this process.

**ACKNOWLEDGMENTS**

Support for this project was provided by a 2017-2018 Connecticut State University – American Association of University Professors (CSU-AAUP) Research Grant and partially supported by National Science Foundation (Grant No. 1422696). We would to like to thank all our respondents for their time in participating in this study and Theresa Parker who assisted with the data analysis for this paper.

**REFERENCES**

Baranuik, C. (2016) Social media loves echo chambers, but the human brain helps create them, *Quartz*.

Barthel, M. Mitchell, A., and Holcomb, J. (2016) Many Americans believe fake news is sowing confusion, *Pew Research*.

Chin, J., Lian, J., Driscoll, K., and Bar, F. (2018) The diffusion of misinformation on social media: Temporal pattern, message, and source, *Computers in Human Behavior*, 83, 278-287.

Endsley, T., Wu, Y., Eep, J., and Reep, J. (2014) The source of the story: Evaluating the credibility of crisis information sources, *Proceedings of the 11th International ISCRAM Conference,* University Park, Pennsylvania, USA, May 2014.

Eriksson, M., and Olsson, E. (2016) Facebook and twitter in crisis communication: A comparative study of crisis communication professionals and citizens, *Journal of Contingencies and Crisis Management,* 24, 4, 198-208. doi:10.1111/1468-5973.12116

Gupta, A., Lamba, H., Kumaraguru, P. and Joshi, A. (2013) Faking sandy: characterizing and identifying fake images on twitter during hurricane sandy, *Proceedings of the 22nd international conference on World Wide Web* (pp. 729-736). ACM.

Halsi, S., Binda, J. and Weirman, S. (2018) It's What Outside That Counts: Finding Credibility Metrics Through Non-Message Related Twitter Features. *Proceedings of the 15th ISCRAM Conference*, Rochester NY USA, May 2018, Kees Boersma and Brian Tomaszewski, eds.

Hiltz, S., Kushma, J., and Plotnick, L. (2014) Use of social media by U.S. public sector emergency managers: Barriers and wish lists. *Proceedings of the 11th International Conference on Information Systems for Crisis Response and Management (ISCRAM)*, University Park, Pennsylvania.

Hosanagar, K., (2016) Blame the Echo Chamber on Facebook. But Blame Yourself, Too. *Wired*, 25 Nov.

Hsieh, G., Chen, J., Mahmud, U.J., and Nichols, J. (2014) You read what you value: understanding personal values and reading interests. *Proceedings of the SIGCHI Conference on Human Factors in Computing Systems* (CHI '14), 983-986.

Hughes, A. L., and Palen, L. (2009) Twitter adoption and use in mass convergence and emergency events, *International Journal of Emergency Management,* 6, 3-4, 248–260.

Imran, M, Castillo, C., Diaz, F. and Vieweg, S. (2015) Processing social media messages in mass emergency: A survey, *ACM Computing Surveys*, 47, 4 Article 67.

Morris, R.M., Counts, S., Roseway, A., Hoff, A., Schwarz, J., (2012) Tweeting is believing? Understanding microblog credibility perceptions. *Proceedings of the ACM 2012 Conference on Computer Supported Cooperative Work*, Seattle, Washington, USA.

Munson, A. S., and Resnick, P. (2010) Presenting diverse political opinions: How and how much. *Proceedings of the SIGCHI Conference on Human Factors in Computing Systems* (CHI '10), 1457-1466.

Ordway D., M., (2017) Fake news and the spread of misinformation, *Journalists Resource*, January 9, 2017.

Palen, L. and Liu, S.B. (2007) Citizen communications in crisis: anticipating a future of ICT-supported public participation. *Proceedings of the Conference on Human Factors in Computing Systems* (CHI), 727-736, San Jose, USA: *ACM Press*.

Pariser, E., (2011) The Filter Bubble: What the Internet Is Hiding from You, Penguin Press, New York.

Pentland, A., (2013) Beyond the echo chamber, *Harvard Business Review*, November Issue.

Plotnick, L. and Hiltz, S.R. (2018) Software innovations to support the use of social media by emergency managers. *International Journal of Human Computer Interaction*, 34, 4, 367-381.

Reuter, C., Backfried, G., Kaufhold, M.A., and Spahr, F. (2018) ISCRAM turns 15: A trend analysis of social media papers 2004- 2017, *Proceedings of the 15th ISCRAM conference*, Rochester NY, May.

Reuter, C., and Kaufhold, M. (2018) Fifteen years of social media in emergencies: A retrospective review and future directions for crisis informatics, *Journal of Contingencies and Crisis Management*, 26, 1, 41-57.

Smith, W.R., Stephens K.K., Robertson, B.W., Li, J., and Murthy, D. (2018) Social media in citizen-led disaster response: Rescuer roles, coordination challenges, and untapped potential, *Proceedings of the 15th ISCRAM conference*, Rochester NY May.

Starbird, K., Maddock, J., Orand, M. Acterman, P. and Mason, R.M. (2014) Rumors, false flags, and digital vigilantes: Misinformation on Twitter after the 2013 Boston Marathon bombing. *Proceedings of*

*Iconference 2014.*

Sutton, J., Palen, L., and Shklovski, I. (2008) Backchannels on the front lines: Emergent uses of social media in the 2007 Southern California wildfires. *Proceedings of the 5th International ISCRAM Conference*, Washington, DC., 624–631.

Wang, Y. and Mark, G. (2017) Engaging with political and social issues on Facebook in college life. *Proceedings of the 2017 ACM Conference on Computer Supported Cooperative Work and Social Computing,* 433-445.

White, J.I., Palen, L., and Anderson, K. (2014). Digital mobilization in disaster response: The work and self-organization of on-line pet advocates in response to Hurricane Sandy, *Proceedings of CSCW'14*, Baltimore, MD, USA.

Wohn, D.Y., and Bowe, B.J. (2016) Micro agenda setters: The effect of social media on young adults' exposure to and attitude toward news. *Social Media and Society*, Jan- March 1-12.

Appendix 1

**The following are a series of general questions about the various social media you use. By social media we mean a software application and associated tools that can be used by groups of individuals to generate and share content and/or engage in peer-to-peer conversation.**

* 10. How often do you visit or use the following? Please click for each row. Notice there is a never option.

|  | Several times a day | About once a day | A few times a week | About once a week | Every few weeks | Less often | Never |
|---|---|---|---|---|---|---|---|
| Facebook | ○ | ○ | ○ | ○ | ○ | ○ | ○ |
| Google+ | ○ | ○ | ○ | ○ | ○ | ○ | ○ |
| Instagram | ○ | ○ | ○ | ○ | ○ | ○ | ○ |
| Snapchat | ○ | ○ | ○ | ○ | ○ | ○ | ○ |
| Twitter | ○ | ○ | ○ | ○ | ○ | ○ | ○ |
| YouTube | ○ | ○ | ○ | ○ | ○ | ○ | ○ |

Other (please specify the social media and frequency of use)